\documentclass[12pt]{book}
\usepackage{plenum}

\catcode`\@=11

\newtoks\@stequation

\def\subequations{\refstepcounter{equation}%
  \edef\@savedequation{\the\c@equation}%
  \@stequation=\expandafter{\theequation}
  \edef\@savedtheequation{\the\@stequation}
  \edef\oldtheequation{\theequation}%
  \setcounter{equation}{0}%
  \def\theequation{\oldtheequation\alph{equation}}}

\def\endsubequations{\setcounter{equation}{\@savedequation}%
  \@stequation=\expandafter{\@savedtheequation}%
  \edef\theequation{\the\@stequation}\global\@ignoretrue
  \vspace*{-12pt} \\}

\catcode`\@=11

\newskip\humongous \humongous=0pt plus 1000pt minus 1000pt
\def\caja{\mathsurround=0pt}
\def\eqalign#1{\,\vcenter{\openup1\jot \caja
	\ialign{\strut \hfil$\displaystyle{##}$&$
	\displaystyle{{}##}$\hfil\crcr#1\crcr}}\,}
\newif\ifdtup

\catcode`@=12
\relax
\def\Q{{\mathchoice
{\setbox0=\hbox{$\displaystyle\rm Q$}\hbox{\raise 0.15\ht0\hbox to0pt
{\kern0.4\wd0\vrule height0.8\ht0\hss}\box0}}
{\setbox0=\hbox{$\textstyle\rm Q$}\hbox{\raise 0.15\ht0\hbox to0pt
{\kern0.4\wd0\vrule height0.8\ht0\hss}\box0}}
{\setbox0=\hbox{$\scriptstyle\rm Q$}\hbox{\raise 0.15\ht0\hbox to0pt
{\kern0.4\wd0\vrule height0.7\ht0\hss}\box0}}
{\setbox0=\hbox{$\scriptscriptstyle\rm Q$}\hbox{\raise 0.15\ht0\hbox to0pt
{\kern0.4\wd0\vrule height0.7\ht0\hss}\box0}}}}
\def\C{{\mathchoice
{\setbox0=\hbox{$\displaystyle\rm C$}\hbox{\hbox to0pt
{\kern0.4\wd0\vrule height0.9\ht0\hss}\box0}}
{\setbox0=\hbox{$\textstyle\rm C$}\hbox{\hbox to0pt
{\kern0.4\wd0\vrule height0.9\ht0\hss}\box0}}
{\setbox0=\hbox{$\scriptstyle\rm C$}\hbox{\hbox to0pt
{\kern0.4\wd0\vrule height0.9\ht0\hss}\box0}}
{\setbox0=\hbox{$\scriptscriptstyle\rm C$}\hbox{\hbox to0pt
{\kern0.4\wd0\vrule height0.9\ht0\hss}\box0}}}}

\def\a{\alpha}

\def\o{\omega}
\def\ho{\hat{\omega}}

\def\bs{\bar{s}}

\def\C{{\cal C}_{g/h}}
\def\cO{{\cal O}}

\def\dif{\partial}

\def\dbar{\bar{\partial}}
\def\del{\nabla}

\def\delh{\hat{\nabla}}

\def\delhp{\hat{\nabla}^+}
\def\delhm{\hat{\nabla}^-}

\def\xx{\hbox{ }^*_*}

\def\ok{\cO (k^{-2} )}

\def\be{\begin{equation}}
\def\ee{\end{equation}}
\def\bs{\begin{subequations}}
\def\es{\end{subequations}}
\def\ben{\begin{enumerate}}
\def\een{\end{enumerate}}

\def\vs{\vskip}

\def\ed{\end{document}}

\def\oa{ {\cal O}(\alpha ')}
\def\oaa{ {\cal O}(\alpha'^2)}

\def\bea{\begin{eqnarray}}
\def\ena{\end{eqnarray}}  \def\eea{\end{eqnarray}}

\begin{document}

\begin{center}
\hfill UCB-PTH-97/41 \\ 
\hfill LBNL-40648 \\  
\hfill hep-th/9708050  \\
\end{center}

\vskip .15in

\noindent
{\bf UNIFICATION OF THE GENERAL NON-LINEAR
SIGMA MODEL\\
AND THE VIRASORO MASTER EQUATION
\footnote{Talk presented by MBH at the NATO Workshop `New Developments
in Quantum Field Theory', June 14-20, 1997, Zakopane, Poland.}}

\vskip 5mm

\author{J. de Boer\footnote{e-mail address: deboer@theorm.lbl.gov}
 and M.B. Halpern\footnote{e-mail address: halpern@theor3.lbl.gov}}

\affiliation{Department of Physics,
University of California at Berkeley\\
366 Le\thinspace Conte Hall, Berkeley, CA 94720-7300, U.S.A.\\
and\\
Theoretical Physics Group, Mail Stop 50A--5101\\
Ernest Orlando Lawrence Berkeley National Laboratory\\
Berkeley, CA 94720, U.S.A.\\}

\vskip .17in

\begin{center} {\bf abstract } \end{center}

\noindent
{\small 
The Virasoro master equation describes a large set of conformal field
theories known as the affine-Virasoro constructions, in the operator
algebra (affine Lie algebra) of the WZW model, while the Einstein
equations of the general non-linear sigma model describe another
large set of conformal field theories. This talk summarizes recent work
which unifies these two sets of conformal field theories, together
with a presumable large class of new conformal field theories. The basic
idea is to consider spin-two operators of the form $L_{ij} \partial x^i
\partial x^j$ in the background of a general sigma model. The requirement
that these operators satisfy the Virasoro algebra leads to a set of
equations called the unified Einstein-Virasoro master equation,
in which the spin-two spacetime field $L_{ij}$ couples to the usual
spacetime fields of the sigma model.
The one-loop form of this unified system is presented, and some
of its algebraic and geometric properties are discussed.}

\section{1. INTRODUCTION}

There have been two broadly 
successful approaches to the construction of conformal
field theories,
\begin{itemize}
\item The general affine-Virasoro construction\refnote{1--7}
\item The general non-linear sigma model\refnote{8--13}
    \hfill (1)
\end{itemize}
\addtocounter{equation}{1}
but, although both approaches have been formulated as Einstein-like
systems\refnote{\cite{sig9,gme}}, the
relation between the two has remained unclear.

This talk summarizes recent work\refnote{\cite{dbh}} which unifies these
two approaches, following the organization of Fig.~1.
The figure shows the two developments (1)  with the
left column (the general affine-Virasoro construction) 
as a special case of the right column (the general non-linear
sigma model). Our goal here is to explain the unification 
shown in the lower right of the figure.

\begin{figure}
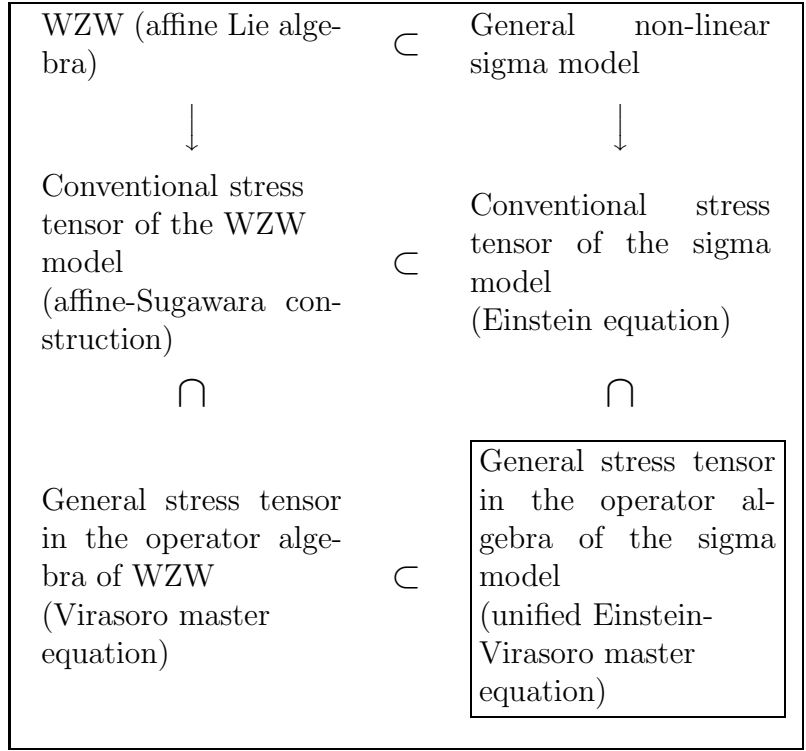

\makebox[3cm]{}
\fbox{
$
\begin{array}{ccc}
\parbox{4cm}{WZW (affine Lie algebra)} 
& \mbox{{\large $\subset$}} & 
\parbox{4cm}{General non-linear sigma model}
\vspace{3mm} \\
\vspace{3mm}
\Big\downarrow & & \Big\downarrow \\
\vspace{3mm}
\parbox{4cm}{Conventional stress\\ tensor 
of the WZW\\ model\\ (affine-Sugawara
  construction)} 
& \makebox[3mm]{} \mbox{{\large $\subset$}} \makebox[3mm]{}  & 
\parbox{4cm}{Conventional stress tensor of the sigma model\\
 (Einstein equation)} \\
\vspace{3mm}
 \mbox{{\Large $\cap$}} & & \mbox{{\Large $\cap$}}  \\
 \vspace{3mm}
 \parbox{4cm}{General stress tensor in the operator
 algebra of WZW\\ 
 (Virasoro master\\
  equation)}
& \mbox{{\large $\subset$}}  & 
\parbox{4cm}{\fbox{\parbox{3.95cm}{General stress tensor in
the operator algebra of the sigma model\\ 
 (unified Einstein-\\
 Virasoro master\\
  equation)}}}
\end{array}
$
}
\begin{center}
\caption{Conformal Field Theory}
\end{center}
\end{figure}

In the general affine-Virasoro construction, a large class of exact 
Virasoro operators\refnote{\cite{hk,rus}}
\be
              T= L^{ab} \xx J_a J_b \xx + iD^a \dif J_a, \qquad
a,b=1 \ldots \dim(g)
\label{avc} \ee
are constructed as quadratic forms in the currents $J$ of the general
affine Lie algebra\refnote{\cite{km,bh}}. The coefficients $L^{ab}=L^{ba}$ and $D^a$ are
called the inverse inertia tensor and the improvement vector
	respectively. The general construction is summarized\refnote{\cite{hk,rus}}
by the (improved)
Virasoro master equation (VME) for $L$ and $D$, and this approach is
the basis of irrational conformal field theory\refnote{\cite{rev}} which includes the
affine-Sugawara\refnote{16--19}
  and coset constructions\refnote{\cite{bh,h1,gko}} as a small subspace. The
construction (\ref{avc}) can also be considered as the 
general Virasoro construction in the operator algebra of the 
WZW model\refnote{\cite{nov,wit}},
which is the field-theoretic realization of the affine algebras.
See Ref.~{\cite{rev}} for a more detailed
history of affine Lie algebra and the affine-Virasoro
constructions.

For each non-linear sigma model, a Virasoro operator\refnote{\cite{sig12}}
\bs
\be
T=-\frac{1}{2\a '} G_{ij} \dif x^i \dif x^j + {\cal O}(\a '^0) =
-\frac{1}{2\a '} G^{ab} \Pi_a \Pi_b + {\cal O}(\a '^0) 
\ee
\be
G^{ab}=e_i{}^a G^{ij} e_j{}^b , \qquad \Pi_a = G_{ab} e_i{}^b \dif x^i, \qquad
i,j,a,b=1,\ldots, \dim(M)
\ee
\label{jjj2} \es
is constructed in a semiclassical expansion on an arbitrary manifold $M$,
where $G_{ij}$ is the metric on $M$ and $G^{ab}$ is the inverse of 
the tangent space metric. 
This is the canonical or conventional
stress tensor of the sigma model and this construction is 
summarized\refnote{\cite{sig9,sig12}} by the Einstein
equations of the sigma model, which couple the metric $G$, the
antisymmetric tensor field $B$ and the dilaton $\Phi$. In what
follows we refer to these equations as the conventional Einstein equations
of the sigma model, to distinguish them from the generalized 
Einstein equations obtained below.

In this paper, we unify these two approaches, using the fact that
the WZW action is a special case of the general sigma model. More
precisely, we study the general Virasoro construction
\bs\be
T=-\frac{1}{\a '} L_{ij} \dif x^i \dif x^j + {\cal O}(\a '^0) =
-\frac{1}{\a '} L^{ab} \Pi_a \Pi_b + {\cal O}(\a '^0) 
\ee\be
i,j,a,b = 1 , \ldots , \dim(M)
\ee \label{jjj3} \es
at one loop in the operator algebra of the general sigma model, where 
$L$ is a symmetric second-rank spacetime tensor field,
the inverse inertia tensor, which is to be determined.
The unified construction is described by a system of equations which we
call
\begin{itemize}
\item the Einstein-Virasoro master equation
\end{itemize}
of the general sigma model. This geometric system, which resides
schematically in the lower right of Fig.~1, describes
the covariant coupling of the spacetime fields
$L$, $G$, $B$ and $\Phi_a$, where the vector field $\Phi_a$ generalizes
the derivative $\del_a \Phi$ of the dilaton $\Phi$.

The unified system contains as special cases the two constructions in
(1): For the particular solution
\be L^{ab}=L_G^{ab} = \frac{G^{ab}}{2} + \oa ,
\qquad \Phi_a = \Phi_a^G = \del_a \Phi   \ee
the general stress tensors (\ref{jjj3}) reduce to the conventional 
stress tensors (\ref{jjj2}) and the Einstein-Virasoro master equation
reduces to the conventional Einstein equations of the sigma model.
Moreover, the unified system reduces to the general affine-Virasoro
construction and the VME when the sigma model is taken to be the WZW
action. In this case we find that the contribution of $\Phi_a$ to
the unified system is precisely the known improvement term of the
VME. 

More generally, the unified system describes a space of conformal
field theories which is presumably much larger than the sum of the
general affine-Virasoro construction and the sigma model with its 
canonical stress tensors.

\section{2. BACKGROUND}

To settle notation and fix concepts which will be important below, we begin
with a brief review of the two known constructions in (1), which are the two
columns of Fig.~1.

\subsection{2.1. The General Affine-Virasoro Construction}

\vs .4cm
\noindent \underline{The improved VME}
\vs .3cm 

The general affine-Virasoro construction, which is the left column of
Figure~1, begins with the currents of a general
affine Lie algebra\refnote{\cite{km,bh}}
\be
J_a(z) J_b(w) = \frac{G_{ab}}{(z-w)^2} + 
\frac{i f_{ab}{}^c J_c(w) }{ z-w} + {\rm reg.}
\label{eq1}
\ee
where $a,b=1\ldots \dim g$ and $f_{ab}{}^c$ are the structure constants
of $g$. For simple $g$, the central term in (\ref{eq1}) has the
form $G_{ab}=k\eta_{ab}$ where $\eta_{ab}$ is the Killing metric of
$g$ and $k$ is the level of the affine algebra. Then the general 
affine-Virasoro construction is\refnote{\cite{hk}}
\be
              T= L^{ab} \xx J_a J_b \xx + iD^a \dif J_a
\label{avc2} \ee
where the coefficients
$L^{ab}=L^{ba}$ and $D^a$ are the inverse inertia tensor and the 
improvement vector respectively. The stress tensor $T$ is a Virasoro
operator
\be
T(z) T(w) = \frac{c/2}{(z-w)^4} + \frac{2 T(w)}{(z-w)^2} + 
\frac{\dif_w T(w)}{z-w} + {\rm reg.} 
\label{ttope}
\ee
iff the improved Virasoro master equation\refnote{\cite{hk}}
\bs
\be
L^{ab} = 2 L^{ac} G_{cd} L^{db} - L^{cd} L^{ef} f_{ce}{}^a f_{df}{}^b
-L^{cd} f_{ce}{}^f f_{df}{}^{(a} L^{b)e} - f_{cd}{}^{(a} L^{b)c} D^d
\label{VME1}
\ee
\be
D^a (2 G_{ab} L^{be} + f_{ab}{}^d L^{bc} f_{cd}{}^e ) = D^e
\label{VME2}
\ee
\be c=2G_{ab} (L^{ab} + 6 D^a D^b) \label{VME3} \ee
\label{VMEall}
\es
is satisfied\footnote{Our convention is $A^{(a} B^{b)} = A^a B^b + A^b B^a$,
 $A^{[a} B^{b]} = A^a B^b - A^b B^a$.} by $L$ and $D$, and
the central charge of the 
construction is given in (\ref{VME3}). The unimproved VME\refnote{\cite{hk,rus}} is 
obtained by setting the improvement vector $D$ to zero.
 
\vs .4cm
\noindent \underline{$K$-conjugation covariance}
\vs .3cm 

A central property of the VME at zero improvement
is $K$-conjugation covariance\refnote{\cite{bh,h1,gko,hk}}
which says that all solutions come in
$K$-conjugate pairs $L$ and $\tilde{L}$,
\bs
\be
L^{ab}+\tilde{L}^{ab} = L_g^{ab}, \qquad
T+\tilde{T} = T_g, \qquad
c+\tilde{c} = c_g 
\ee\be
T(z)\tilde{T}(w) = {\rm reg.}
\ee
\es
whose $K$-conjugate stress tensors $T,\tilde{T}$ commute and add
to the affine-Sugawara construction [15--18] on $g$
\be
T_g = L_g^{ab} \xx J_a J_b \xx .
\label{affsug}
\ee
For simple $g$, the inverse inertia tensor of the
affine-Sugawara construction is
\be
L_g^{ab} = \frac{\eta^{ab}}{2k+Q_g} = \frac{\eta^{ab}}{2k} + \ok = 
\frac{G^{ab}}{2} + \ok 
\label{affa}
\ee
where $\eta^{ab}$ is the inverse Killing metric of $g$ and $Q_g$ is the 
quadratic Casimir of the adjoint. $K$-conjugation covariance can be used
to generate new solutions $\tilde{L}=L_g-L$ from old solutions $L$ and
the simplest application of the covariance generates 
the coset constructions\refnote{\cite{bh,h1,gko}} as $\tilde{L}=L_g-L_h=L_{g/h}$.

\vs .4cm
\noindent \underline{Semiclassical expansion}
\vs .3cm 

At zero improvement, the high-level or semiclassical expansion\refnote{\cite{hl,rev}}
 of
the VME has been studied in some detail. On simple $g$, the leading term
in the expansion has the form
\bs\be
L^{ab} = \frac{P^{ab}}{2k} + \ok, \qquad c = {\rm rank}(P) + 
{\cal O}(k^{-1}) 
\ee\be
P^{ac} \eta_{cd} P^{db} = P^{ab}
\ee
\label{expan1}
\es
where $P$ is the high-level projector of the $L$ theory. These are the 
solutions of the classical limit of the VME,
\be L^{ab} = 2 L^{ac} G_{cd} L^{db} + \ok 
\label{expan2}
\ee
but a semiclassical quantization condition\refnote{\cite{hl}} provides a 
restriction on the allowed projectors. In the partial classification
of the space of solutions by graph theory\refnote{\cite{gt,ggt,rev}}, the projectors 
$P$ are closely related to the
adjacency matrices of the graphs.

\vs .4cm
\noindent \underline{Irrational conformal field theory}
\vs .3cm 

Given also a set of antiholomorphic currents $\bar{J}_a$, $a=1\ldots
\dim(g)$, there is a corresponding antiholomorphic Virasoro construction
\be
\bar{T}= L^{ab} \xx \bar{J}_a \bar{J}_b \xx + iD^a \dif \bar{J}_a
\label{avc3} \ee
with $\bar{c}=c$. Each pair of stress tensors $T$ and $\bar{T}$ then
defines a conformal field theory (CFT) labelled by $L$ and $D$. 
Starting from the modules of affine $g \times g$, the 
Hilbert space
of a particular CFT is obtained\refnote{\cite{hy,rp,rev}} by modding out by the
local symmetry of the Hamiltonian. 

It is known that the CFTs of the master equation have generically
irrational central charge, even when attention is restricted to the
space of unitary theories, and the study of all the CFTs of the 
master equation is called irrational conformal field theory (ICFT),
which contains the affine-Sugawara and coset constructions as a 
small subspace. 

In ICFT at zero improvement, 
world-sheet actions are known for the following cases: the 
	affine-Sugawara constructions (WZW models\refnote{\cite{nov,wit}}),
the coset constructions (spin-one gauged WZW models\refnote{\cite{br}}) and the
generic ICFT (spin-two gauged WZW models\refnote{\cite{hy,ts3,bch}}). The spin-two
gauge symmetry of the generic ICFT is a consequence of
$K$-conjugation covariance.

See Ref.~\cite{rev} for a comprehensive review of ICFT, 
and Ref.~\cite{blocks} for a 
recent construction of a set of
semiclassical blocks and correlators in ICFT.

In this talk, we restrict ourselves to holomorphic stress tensors, and
the reader is referred to Ref.~\cite{dbh} for the antiholomorphic version.

\vs .4cm
\noindent \underline{WZW model}
\vs .3cm 

The left column of Fig.~1 can be considered as the set of 
constructions in the operator algebra of the WZW model,
which is affine Lie algebra.

The WZW action is a special case of the general nonlinear sigma model, 
where the target space is a group manifold $G$ and $g$ is the 
algebra of $G$. 

\subsection{2.2. The General Non-Linear Sigma Model}

The general non-linear sigma model (the right column 
of Fig.~1) has been extensively 
studied\refnote{\cite{sig1,sig1a,sig2,sig3,sig4,sig5,sig6,sig7,sig8,sig9,sig11,sig10,sig12,sig13}}.

The Euclidean action of the general non-linear sigma model is
\bs\be
S=\frac{1}{2\a '} \int d^2 z  (G_{ij} + B_{ij}) \dif x^i \dbar x^j
\label{sigmaact}
\ee\be
d^2z=\frac{dxdy}{\pi}, \qquad z=x+iy, \qquad
H_{ijk} = \dif_i B_{jk} + \dif_j B_{ki} + \dif_k B_{ij}.
\ee\es
Here $x^i$, $i=1\ldots \dim(M)$ are coordinates with the dimension
of length on a general manifold $M$ and $\a'$, with dimension length
squared, is the string tension or Regge slope. 
The fields $G_{ij}$ and $B_{ij}$ are 
the (covariantly constant) metric and antisymmetric tensor field on $M$.

We also introduce a covariantly constant vielbein $e_i{}^a$, $a=1\ldots
\dim(M)$ on $M$ and use it to translate between Einstein and 
tangent-space indices, e.g.
$G_{ij}=e_i{}^a G_{ab} e_j{}^b$, 
where $G_{ab}$ is the covariantly constant metric on tangent space. Covariant
derivatives are defined as usual in terms of the spin connection,
$R_{ija}{}^b$ is the
Riemann tensor and $R_{ab}=R_{acb}{}^c$ is the Ricci tensor.
It will also be convenient to define the generalized connections
and covariant derivatives with torsion,
\bs\be
\delh^{\pm}_i v_a = \dif_i v_a - \hat{\o}^{\pm}_{ia}{}^b v_b
\label{tordefa}
\ee\be
\ho^{\pm}_{ia}{}^b=\o_{ia}{}^b \pm \frac{1}{2} H_{ia}{}^b
\label{tordefb}
\ee \be
\hat{R}^{\pm}_{ija}{}^b= (\dif_i \ho^{\pm}_j-\dif_j \ho^{\pm}_i
-[\ho_i^{\pm},\ho_j^{\pm}])_a{}^b
\label{tordefd}
\ee
\label{tordef}
\es
where $\ho^{\pm}_{iab}$ is antisymmetric under $(a,b)$ interchange and
$\hat{R}^{\pm}_{ijab}$ is pairwise antisymmetric in $(i,j)$ and $(a,b)$.

Following Banks, Nemeschansky and Sen\refnote{\cite{sig12}}, the 
canonical or
conventional stress tensors
of the general sigma model have the form
\bs\be
T_G = -\frac{G_{ij}}{2\a '} \dif x^i \dif x^j + \dif^2 \Phi + T_1
+ {\cal O}(\a ')
\ee\be
\qquad  = -\frac{G^{ab}}{2\a '} \Pi_a \Pi_b + \dif^2 \Phi + T_1
+ {\cal O}(\a ')
\ee \be
\Pi_a=G_{ab} e_i{}^b \dif x^i, \qquad \bar{\Pi}_a=G_{ab} e_i{}^b
\dbar x^i
\label{defpi}
\ee \label{jjj19} \es
where $\Phi$ is the dilaton and $T_1$ is a finite one-loop 
counterterm which depends on the renormalization scheme.
The condition that $T_G$ is one-loop conformal
reads\refnote{\cite{sig9}}
\bs\be
R_{ij}+\frac{1}{4} (H^2)_{ij} - 2 \del_i \del_j \Phi = 
{\cal O}(\a ') 
\label{conf1}
\ee\be
\del^k H_{kij} - 2 \del^k \Phi H_{kij} = {\cal O}(\a')
\label{conf2}
\ee\be
c_G=\bar{c}_G = \dim(M) + 3 \alpha' 
(4 |\del \Phi|^2 - 4 \del^2 \Phi + R + \frac{1}{12} H^2) + 
{\cal O}(\a '^2)
\label{conf3}
\ee
\label{conf}
\es
where (\ref{conf1}) and (\ref{conf2}) are the conventional
Einstein equations of 
the sigma model and (\ref{conf3}) is the central charge of the
construction. The result for the central charge includes two-loop 
information, but covariant constancy of the field-dependent part
of the central charge follows by Bianchi identities from the
Einstein equations, so all three relations 
in (\ref{conf})
can be obtained with a little
thought from the one-loop calculation. It will also be useful to note that
the conventional
Einstein equations (\ref{conf1}),(\ref{conf2}) can be written in
either of two equivalent forms
\be
\hat{R}^{\pm}_{ij} - 2 \delh^{\pm}_i \delh^{\pm}_j \Phi = 
{\cal O}(\a ')
\label{conf4}
\ee
by using the generalized quantities (\ref{tordef}) with torsion.

\vs .4cm
\noindent \underline{WZW data}
\vs .3cm 

The WZW action is a special case of the general sigma
model (\ref{sigmaact}) on a group manifold $G$. Identifying the
vielbein $e$ on $M$ with the left-invariant vielbein $e$
on $G$, we find that $J_a=\frac{i}{\sqrt{\alpha '}} \Pi_a$
are the classical currents of WZW and 
\be
G_{ab} = k \eta_{ab}, \qquad
H_{ab}{}^c = \frac{1}{\sqrt{\a'}} f_{ab}{}^c.
\label{wzwdata}
\ee
Here
$f_{ab}{}^c$ and $\eta_{ab}$ are the structure constants and the Killing
metric of $g$ and $k$ is the level of the affine algebra. From this data,
one also computes
\bs\be
\o_{ab}{}^c = -\frac{1}{2\sqrt{\a'}} f_{ab}{}^c 
\ee\be
\ho^+_{ab}{}^c = 0, \qquad \ho^-_{ab}{}^c = -\frac{1}{\sqrt{\a'}} f_{ab}{}^c
 \label{auxn17} \ee
\be
\hat{R}^{\pm}_{ija}{}^b = 0.
\label{wzwrhat}
\ee\es
Manifolds with 
vanishing generalized Riemann tensors are 
called parallelizable\refnote{\cite{sig6,sig11}}.

\subsection{2.3. Strategy}

As seen in Fig.~1, 
our strategy here is a straightforward generalization of the VME
to the sigma model, following the relation of the general 
affine-Virasoro construction to the WZW model. In the operator algebra
of the general sigma model, we use the technique of Banks et
al.\refnote{\cite{sig12}} to study the general Virasoro construction
\bs\be
T = -\frac{L_{ij}}{\a '} \dif x^i \dif x^j 
+ {\cal O}(\a '^0)
 = -\frac{L^{ab}}{\a '} \Pi_a \Pi_b 
+ {\cal O}(\a '^0)
\label{genvira}
\ee \be
\dbar T = 0
\label{genvirc}
\ee\be
<T(z)T(w)> = 
\frac{c/2}{(z-w)^4} + \frac{2<T(w)>}{(z-w)^2} 
 + \frac{<\dif T(w)>}{(z-w)} + {\rm reg.}
\label{genvird}
\ee \label{genvir} 
\es
where the dilatonic contribution is included at 
${\cal O}(\a '^0)$
 and $L$ is a symmetric
second-rank spacetime tensor field (the inverse inertia tensor)
to be determined.

It is clear that this one-loop construction includes the conventional stress tensor
$T_G$ of the general sigma model, as well as the general
affine-Virasoro construction when the sigma model is chosen to be
WZW.

\section{3. CLASSICAL PREVIEW OF THE CONSTRUCTION}

The classical limit of the general construction (\ref{genvira})
can be studied with the classical equations of motion of the general
sigma model
\be
\dbar \Pi_a + \bar{\Pi}_b \Pi_c \ho^{+bc}{}_a = 0
\label{eqmot}
\ee
where $\Pi,\bar{\Pi}$ are defined in (\ref{defpi})  and $\ho^{\pm}$
are the generalized connections (\ref{tordefb}) with torsion.

One then finds that 
the classical stress tensor is holomorphic
\be
T=-\frac{L^{ab}}{\a'} \Pi_a \Pi_b, \qquad
\dbar T = 0
\label{tcla}
\ee
iff the inverse inertia tensor is covariantly constant
\be
\delhp_i L^{ab} = 0
\label{lconst}
\ee
where $\delh^{\pm}$ are the generalized covariant derivatives (\ref{tordefa})
with torsion. Further discussion of this covariant-constancy condition
is found in Sections~5.2 and especially~5.5, which places the relation
in a more geometric context.

To study the classical Virasoro conditions, we introduce Poisson brackets
in Minkowski space, and study the classical chiral stress tensor
\be
T_{++}= \frac{1}{8\pi \a'} L^{ab} J^+_a J^+_b
\ee
where $J^+_a$ is the Minkowski-space version of $\Pi_a$.
This stress tensor satisfies the equal-time Virasoro algebra iff
\be
L^{ab}= 2 L^{ac} G_{cd} L^{db},
\label{lconst2}
\ee
which is the analogue on general manifolds of the high-level
or classical limit (\ref{expan2}) of the VME on group manifolds.

\section{4. THE UNIFIED EINSTEIN-VIRASORO MASTER EQUATION}

We summarize here the results obtained by enforcing the Virasoro
condition (\ref{genvird}) at one loop. 
Details of the relevant background field
expansions, Feynman diagrams and dimensional regularization
can be found in Ref.~{\cite{dbh}}.

Including the one-loop dilatonic and counterterm contributions,
the holomorphic stress tensor $T$ is
\bs\be
T = - L^{ab} (\frac{\Pi_a \Pi_b}{\a '} + \frac{1}{2} \Pi_c \Pi_d 
H_{ae}{}^c H_b{}^{ed} ) + \dif (\Pi_a \Phi^a) + {\cal O}(\a')
\label{countera}
\ee
\vskip -.15in
\be
a,b = 1, \ldots, \dim(M)
\label{counterd} \ee
\label{counter}
\es
where $L^{ab}=L^{ba}$ is the inverse inertia
tensor and $\Pi_a$ is defined in (\ref{defpi}).
The second term in $T$ is a finite one-loop counterterm
which characterizes our renormalization scheme. 
The quantity $\Phi^a$ in 
(\ref{countera})
is called the dilaton vector, and we will see below that the dilaton
vector includes the conventional dilaton as a special case.

The necessary and sufficient 
condition that $T$ satisfies the Virasoro algebra is
the unified Einstein-Virasoro
master equation
\bs\be
L^{cd} \hat{R}^+_{acdb} + \delhp_a \Phi_b = {\cal O}(\a')
\label{eva}
\ee\be
\Phi_a = 2 L_a{}^b \Phi_b + \oa
\label{evn}
\ee\be
\delhp_i L^{ab} = {\cal O}(\a') 
\label{evb}
\ee\be
\eqalign{ L^{ab}  = & 2 L^{ac} G_{cd} L^{db} \cr
 & - \a' (L^{cd} L^{ef} H_{ce}{}^a H_{df}{}^b + 
 L^{cd} H_{ce}{}^f H_{df}^{(a} L^{b)e} ) \cr
& - \a' (L^{c(a} G^{b)d} \del_{[c} \Phi_{d]} ) + {\cal O}(\a'^2) \cr}
\label{evc}
\ee\be
c = 2 G_{ab} L^{ab} + 6 \a' (2 \Phi_a \Phi^a - \del_a \Phi^a)
 + {\cal O}(\a'^2)
\label{evd}
\ee
\label{ev}
\es
where the first line of (\ref{evc}) is the classical master equation in 
(\ref{lconst2}).

In what follows, we refer to (\ref{eva}) as the 
generalized Einstein equation of the sigma model, and equation
(\ref{evn}) is called the eigenvalue relation
of the dilaton vector.
Equation (\ref{evc})
is called the generalized Virasoro master equation
(VME) of the sigma model. 
The central charge (\ref{evd})
is consistent\refnote{\cite{dbh}} by Bianchi identities with the rest of the unified system.
The ${\cal O}(\a')$ corrections to the covariant-constancy
condition (\ref{evb}) can be computed in principle
from the solutions of the generalized VME.  

\vs .4cm
\noindent \underline{Some simple observations}
\vs .3cm 

\noindent
1. Algebraic form of the generalizedVME.
In parallel
with the VME, the generalized VME (\ref{evc}) is an algebraic 
equation for $L$. This follows because any derivative of 
$L$ can be removed by using the
covariant-constancy condition (\ref{evb}). 


\vs .3cm

\noindent
2. Semiclassical solutions of the generalized VME. The solutions
of (\ref{evb}) and (\ref{evc}) have the form
\bs\be
L_a{}^b  = \frac{1}{2} P_a{}^b + \oa
\ee\be
\hat{\nabla}_i^+ P_a{}^b = 0
\label{pqr}
\ee\es
where $P$ is a covariantly constant projector,
in parallel with the form (\ref{expan1}) of the high-level
solutions of the VME. The solutions of (\ref{pqr})
are further discussed in Section~5.5.

\vs .3cm

\noindent
3. Correspondence with the VME. The non-dilatonic terms of the
generalized
VME (\ref{evc}) have exactly the form of the 
unimproved VME
(see eq. (\ref{VME1})), after the covariant substitution
\be
f_{ab}{}^c \rightarrow \sqrt{\a'} H_{ab}{}^c
\label{auxn11}
\ee
for the general sigma model. This correspondence is the inverse of the
WZW datum in (\ref{wzwdata}),
\be H_{ab}{}^c = \frac{1}{\sqrt{\a'}} f_{ab}{}^c
\ee
which means that, for 
the special case of 
WZW, the non-dilatonic terms of the
generalized VME will reduce correctly to those of the 
unimproved VME. We return
to complete the WZW reduction in Section~5.2.

\vs .3cm

\noindent
4. Dilaton solution for the dilaton vector. According to the classical
limit (\ref{lconst2})
of the generalized VME, one solution of the eigenvalue relation 
(\ref{evn}) for the dilaton vectors is
\be
 \Phi^a(\Phi)  \equiv 2 L^{ab} \del_b \Phi 
\label{eig}
\ee 
In what follows, this solution is called the dilaton solution, and we
shall see in the following section that the scalar field $\Phi$ is in
fact the conventional dilaton of the sigma model.

\section{5. PROPERTIES OF THE UNIFIED SYSTEM}

\subsection{5.1. The Conventional Stress Tensors of the Sigma Model}

In this section, we check that the conventional stress tensors of the
sigma model are correctly included in the unified system.

In the full system, the conventional stress tensor
$T_G$ of the sigma
model corresponds to the particular solution of the generalized VME
whose classical limit is 
\be
L^{ab} = L_G^{ab} = \frac{G^{ab}}{2} + \oa
\label{norms}
\ee
where $G^{ab}$ is the inverse of the metric in the sigma model action. 
The covariant-constancy condition (\ref{evb}) is 
trivially solved to this order because $\delh^{\pm}_i G_{ab}=0$.

To obtain the form of $T_G$ through one loop, 
we must also take
the dilaton solution (\ref{eig}) for the dilaton vector, so that the
dilaton contributes to the system as
\be
\Phi_a =\Phi_a^G = \del_a \Phi + \oa,\qquad
\label{dnorma}
\del_{[a} \Phi_{b]}^G = \oa .
\ee
The relations (\ref{norms}) and (\ref{dnorma}) then tell us that the 
generalized Einstein equation (\ref{eva})
simplifies to the conventional Einstein equation
\be
\hat{R}^{\pm}_{ab} - 2 \delh^{\pm}_a \delh^{\pm}_b \Phi = 
{\cal O}(\a ').
\ee
Moreover, eq. (\ref{dnorma}) tells us that the dilaton terms
do not contribute to the generalized VME in this case, and we may
easily obtain
\bs\be
L^{ab}=L^{ab}_G = \frac{G^{ab}}{2} -
\frac{\alpha'}{4} (H^2)^{ab} + \oaa 
\label{deflg}
\ee\be
\delh^{\pm}_i L^{ab}_G = -\frac{\alpha'}{4} 
 \delh^{\pm}_i (H^2)^{ab}  + \oaa
\ee\be
T_G(\Phi) = -\frac{G^{ab}}{2\a'}\Pi_a \Pi_b  + \dif^2 \Phi + \oa
\label{zz1}
\ee \es
by solving the generalized VME through the indicated order. In this
case, the stress tensor counterterm in (\ref{countera})
cancels against the $\oa$ correction to $L_G$,
and (\ref{zz1}) are consistent with (\ref{jjj19}).
In what follows, the stress
tensor $T_G(\Phi)$ is
called the conventional stress
tensor of the sigma model.

To complete the check, we evaluate the central charge
$c=c_G(\Phi)$ in this case,
\bs\be
c_G(\Phi)=2 G_{ab} (\frac{G^{ab}}{2} -\frac{\a'}{4}  (H^2)^{ab} ) 
+ 6 \a' (2 |\del \Phi|^2 - \del^2 \Phi) + \oaa 
\label{ccca}
\ee\be
\qquad = \dim(M)  + 3 \a' (4 |\del \Phi|^2 - 2 \del^2 \Phi -
\frac{1}{6} H^2 ) + \oaa
\label{cccb}
\ee\be
\qquad = \dim(M)  + 3 \a' (4 |\del \Phi|^2 - 4 \del^2 \Phi +R +
\frac{1}{12} H^2 ) + \oaa
\label{cccc}
\ee \label{ccc} \es
which agrees with the conventional central charge in (\ref{conf3}). To
obtain the usual form in (\ref{cccc}), we used the conventional 
Einstein equations (\ref{conf1}) in the form $R=2\del^2 \Phi -\frac{1}{4}
H^2$.

We also note the form of the system for
$L=L_G$ with general dilaton vector $\Phi^G_a$,
\bs\be
T_G(\Phi_a) = -\frac{G^{ab}}{2\a'} \Pi_a \Pi_b + \dif(\Pi^a \Phi_a^G) +\oa
\label{qqa}
\ee\be
c_G(\Phi_a) = \dim(M) + 3\a' (4 \Phi^G_a \Phi_G^a - 4 \del_a \Phi^a_G + R
 + \frac{1}{12} H^2 ) + \oaa
\label{qqb}
\ee\be
\hat{R}^+_{ab} - 2 \delhp_a \Phi^G_b = \oa
\label{qqc}
\ee \label{qq} \es
where $\Phi_a^G$ is unrestricted because its eigenvalue equation is trivial.

\subsection{5.2. WZW and the Improved VME}

In this section we check that, for the special case of WZW, the unified
system reduces to the improved
VME (\ref{VME1}), where the improvement vector $D$ is constructed
from the general dilaton vector.

Using the WZW datum above we find that the
generalized VME (\ref{evc}) has the
form
\be
L^{ab}   =   (\mbox{{\rm usual $L^2$ and $L^2f^2$ terms}}) + 
\sqrt{\a'} f_{cd}{}^{(a} L^{b)c} \Phi^d + \oaa
\ee
when the sigma model is taken as WZW. 
The terms in parentheses are the usual terms
(see eq. (\ref{VME1})) of the unimproved VME.
Next, we solve the generalized Einstein
equation (\ref{eva}) to find (using $\hat{R}^{\pm}=0$)
that the dilaton vector is a constant
\be
\dif_i \Phi^a = 0. 
\label{consa}
\ee
It follows that the dilaton vector
can be identified 
with the improvement vector of the VME
in (\ref{VME1})
\be
D^a \equiv - \sqrt{\a'} \Phi^a = {\rm constant}.
\label{consb}
\ee
Moreover, the solution of 
the covariant-constancy condition (\ref{evb}) is
\be
\dif_i L^{ab}= 0, \qquad 
L^{ab} = {\rm constant}
\ee
because $\hat{\omega}^+=0$. 
This completes the 
recovery of the improved VME in
(\ref{VME1}).

The central charge reduces in this case
to
\be
c=2 G_{ab} (L^{ab} + 6 D^a D^b) + {\cal O}(k^{-2})
\label{redca}
\ee
in agreement with the central charge (\ref{VME3}) of the improved VME.
We finally note that the eigenvalue relation (\ref{evn})
of the dilaton
vector can be rewritten with (\ref{consb}) as
\be
2 L^{ab} G_{bc} D^c = D^a + {\cal O}(k^{-2})
\label{eeig}
\ee
which is recognized as 
the leading term of the exact eigenvalue
relation (\ref{VME2}) of the improved VME.
This completes the one-loop check of the unified
Einstein-Virasoro master equation
 against the 
improved VME.

\subsection{5.3. Alternate Forms of the Central Charge}

Using the generalized Einstein equation and the generalized VME,
the central charge (\ref{evd}) can
be written in a variety of forms,
\bs\be
c = 2 L_a{}^a + 6 \a' (2 \Phi_a \Phi^a - 
\delhp_a \Phi^a) + \oaa
\label{varca} \ee\be
= 4 L_a{}^b L_b{}^a + 2 \a' 
\left[ L_b{}^e L_d{}^f H^{bda} H_{efa} 
+3(2 \Phi_a \Phi^a - \delhp_a \Phi^a) \right] + \oaa
\label{varcc} \ee\be
= {\rm rank}(P) + 2 \a' 
\left[ L_a{}^b L_c{}^e (4L_d{}^f - 3 \delta_d{}^f ) H^{acd} H_{bef} 
+3(2 \Phi_a \Phi^a - \delhp_a \Phi^a) \right] + \oaa
\label{varcd} \ee\be
= {\rm rank}(P) + 2 \a' 
\left[  3 L^{ab} \hat{R}^+_{ab} + 
L_a{}^b L_c{}^e (4L_d{}^f - 3 \delta_d{}^f ) H^{acd} H_{bef} 
+6( \Phi_a \Phi^a - \delhp_a \Phi^a) \right] + \oaa
\label{varce} \ee\be
= 4 L_a{}^b L_b{}^a + 2 \a' 
\left[  3 L^{ab} \hat{R}^+_{ab} + 
L_a{}^b L_c{}^e  H^{acd} H_{bed} 
+6( \Phi_a \Phi^a - \delhp_a \Phi^a) \right] + \oaa.
\label{varcf} \ee
\label{varc} \es
The first form in (\ref{varca}) is the `affine-Virasoro form' of
the central charge.
The form in (\ref{varce}), with the first occurence
of the generalized Ricci tensor, is called the `conventional form'
of the central charge because it reduces easily to the central charge
of the conventional stress tensor
\be
c_G(\Phi) =\dim (M) + 3 \a' (4 | \del \Phi |^2 - 4 \del^2 \Phi + R + 
\frac{1}{12} H^2) + \oaa
\ee
when $P=G$ and $\Phi_a^G=\del_a \Phi$. The conventional form is also
the form in which
we found\refnote{\cite{dbh}} it most convenient
to prove the constancy of $c$ 
\be
\partial_i c = \hat{\nabla}^+_i c = \oaa
\ee
using the Bianchi identities and the rest of the unified system.
The final form of $c$ in (\ref{varcf}) is the form which we believe
comes out directly from the two-loop computation.

\subsection{5.4. Solution Classes and a Simplification}

\vs .4cm
\noindent \underline{Class I and Class II solutions}
\vs .3cm 

The solutions of the unified system (\ref{ev}) can be divided into
two classes:

\vs .1cm

\noindent
Class I. $T$ conformal but $T_G(\Phi_a)$ not conformal
\vs .1cm

\noindent
Class II. $T$ and $T_G(\Phi_a)$ both conformal.
\vs .1cm

\noindent
The 
distinction here is based on whether or not (in addition
to the generalized Einstein equation) 
the dilaton-vector Einstein
equation in (\ref{qqa}) 
is {\it also} satisfied. In the case when
the dilaton solution $\Phi_a(\Phi)$ in
(\ref{eig}) is taken for the dilaton vector,
the question is whether or not the background sigma model is itself
conformal in the conventional sense.

In Class I, we are constructing a conformal stress tensor $T$ in the
operator algebra of a sigma model whose conventional stress tensor 
$T_G(\Phi_a)$ with general dilaton vector is not conformal.
This is a situation not encountered in the general
affine-Virasoro construction because the conventional stress-tensor 
$T_g$ of the WZW model is the affine-Sugawara construction,
which is conformal. It is expected that Class I solutions are generic
in the unified system, since there are so many non-conformal sigma
models, but there are so far no non-trivial\footnote{Trivial
examples in Class I
are easily constructed as tensor products of conformal and
non-conformal theories.}
examples (see however Ref.~{\cite{chun}},
which proposes a large set of candidates).

In Class II, we are constructing a conformal stress tensor $T$ in
the operator algebra of a sigma model whose conventional stress
tensor $T_G(\Phi_a)$ with general dilaton vector
is conformal. This class includes the 
case where the conventional stress tensors $T_G(\Phi)$ are conformal
so that the sigma model is conformal in the conventional sense.
The general affine-Virasoro
construction provides a large set of non-trivial examples in Class II
when the sigma model is the WZW action.
Other examples are known from the general affine-Virasoro construction
which are based on coset constructions, 
instead of WZW. In particular, Halpern et al.\refnote{\cite{lie}}
 construct exact Virasoro
operators in the Hilbert space of a certain class of
$g/h$ coset constructions,
and we are presently studying these conformal field theories
as Class II solutions in the
sigma model description of the coset constructions (see also
the Conclusions).

It is also useful to subdivide Class II solutions into Class IIa and IIb.
In Class IIb, we require the natural identification
\be
\Phi_a = 2 L_a{}^b \Phi_b^G + \oa
\label{aabb}
\ee
which solves (\ref{evn}), and Class IIa is the set of solutions in Class II
without this identification. Note in particular that Class IIb contains
all solutions in Class II with the dilaton solution $\Phi_a(\Phi)$ in
(\ref{eig}).

\vs .4cm
\noindent \underline{Simplification for Class IIb with the dilaton solution}
\vs .3cm 

A simplification in Class IIb follows for the dilaton solution
$\Phi_a(\Phi)$. In this case the unified system reads
\bs\be
R_{ij}+\frac{1}{4} (H^2)_{ij} - 2 \del_i \del_j \Phi = 
{\cal O}(\a ') 
\label{xyza}
\ee\be
\del^k H_{kij} - 2 \del^k \Phi H_{kij} = {\cal O}(\a')
\label{xyzb}
\ee\be
\delhp_i L^{ab} = {\cal O}(\a') , \qquad
\delhm_i \bar{L}^{ab} = {\cal O}(\a') 
\label{xyzc}
\ee\be
\eqalign{ L^{ab}  = & 2 L^{ac} G_{cd} L^{db} \cr
 & - \a' (L^{cd} L^{ef} H_{ce}{}^a H_{df}{}^b + 
 L^{cd} H_{ce}{}^f H_{df}^{(a} L^{b)e} ) \cr
& - \a' (L^{c(a} G^{b)d} \del_{[c} \Phi_{d]} ) + {\cal O}(\a'^2) \cr}
\label{xyzd}
\ee \be
c = 2 G_{ab} L^{ab} + 6 \a' (2 \Phi_a \Phi^a - \del_a \Phi^a)
 + {\cal O}(\a'^2)
\label{xyzf}
\ee \be
\Phi_a=\Phi_a(\Phi)=2 L_a{}^b \del_b \Phi.
\label{xyzh}
\ee
\label{xyz} \es
This simplified system is close in spirit to the VME of the
general affine-Virasoro construction: The solution of the conventional
Einstein equation in (\ref{xyza}), (\ref{xyzb})
provides a conformal background, in which we need
only look for solutions of the generalized VME in the form
\be
L_a{}^{b} = \frac{P_a{}^{b}}{2} + \oa
\ee
where $P$ is a covariantly constant projector.
Moreover, as in the VME, it has been shown\refnote{\cite{dbh}}
that all solutions of the
simplified system (\ref{xyz})
exhibit $K$-conjugation covariance, so that
\be
\tilde{T}\equiv T_G-T, \qquad \tilde{c}=c_G-c
\label{covprop}
\ee
is also a conformal stress tensor when $T$ is conformal.

\subsection{5.5. Integrability Conditions}

The inverse inertia tensor $L^{ab}$ is a
second-rank symmetric spacetime 
tensor, and we know that its associated projector
$P$ is covariantly constant
\be
\delhp_i P_a{}^b = 0
\label{pcon}
\ee
Operating with a second covariant derivative, we find that the 
integrability conditions
\be
\hat{R}^+_{cd}{}^{ae} P_e{}^b + \hat{R}^+_{cd}{}^{be} P_e{}^a = 0
\label{aux9}
\ee
follow as necessary conditions for the existence of solutions to (\ref{pcon}).

On any manifold, there is always at least one solution to the
covariant-constancy condition (\ref{pcon}) and 
its integrability conditions (\ref{aux9}), namely
\bs\be
P^{ab}=G^{ab}
\ee\be
\hat{R}^{\pm ab}_{cd} +\hat{R}^{\pm ba}_{cd} = 0 
\ee\be
L^{ab}=L_G^{ab} = 
\frac{G^{ab}}{2} + \oa
\ee\es
where $G_{ab}$ is the metric of the sigma model action. This solution 
corresponds to the classical limit of the conventional sigma model stress tensor,
as discussed in Section~5.1. For WZW, the integrability conditions 
(\ref{aux9}) are also trivially satisfied (because 
$\hat{R}^{\pm}_{abcd}=0$) and the general solutions of the
covariant-constancy conditions were obtained for this case in
Section~5.2.

In general we are interested in the classification of manifolds with at
least one more solution $P^{ab}$, 
beyond $G^{ab}$. In what follows, we outline the sufficient and 
necessary condition for this phenomenon.

In a suitable basis, any projector $P$ can be
written as
\be \label{aaa1}
P=\left(  \begin{array}{c|c} {\bf 1} & 0 \\ \hline 
   0 & 0 \end{array} \right) .
\ee
Inserting this form in (\ref{pcon}) and (\ref{aux9}) shows then that
$\hat{R}^+$ and $\hat{\omega}^+$ must be `block diagonal'
in the same basis, i.e. they can be written as
\be
(\hat{R}^+_{cd})_a^b = \left(  \begin{array}{c|c} A_{cd} & 0  \\ \hline 
   0 & D_{cd}  \end{array} \right) , \qquad
(\hat{\omega}^+_{i})_a^b = \left(  \begin{array}{c|c} D_i & 0 \\ \hline
   0 & E_i  \end{array} \right)
\ee
for some matrices $A_{cd},B_{cd},D_i,E_i$. Thus, a necessary
condition for new solutions to the covariant-constancy condition to
exist is that $\hat{R}$ and $\hat{\omega}$ should be block diagonal.

Conversely, given a block diagonal $\hat{\omega}^+$, we can construct 
a new solution to the covariant-constancy condition with $P$ given
in (\ref{aaa1}). In fact, with $\hat{\omega}^+$ written in terms of
the smallest possible blocks we can classify all possible solutions to
the covariant constancy condition. If we denote the smallest 
diagonal blocks of $\hat{\omega}^+$ by $D_1,\ldots,D_k$, then
the most general covariantly constant projector is
\be P= p_1 {\bf 1}_1 + \ldots + p_k {\bf 1}_k
\ee
where $p_i \in \{0,1\}$ and ${\bf 1}_j$ is the matrix which consists of
the identity matrix in the $j^{\rm th}$ block and zeroes everywhere
else. In the case when one of the blocks in $\hat{\omega}^+$ is
zero, say $D_j$, then $p_j {\bf 1}_j$ can be replaced by
an arbitrary projector $P_j$ in the $j^{\rm th}$ subspace.

New solutions obtained following this procedure are discussed
in the Conclusions. 

Mathematically, the problem of finding block-diagonal curvatures
is the problem of finding manifolds with reducible holonomy. 
In the absense of torsion, it is known that block-diagonal curvatures exist
only on product manifolds, but in the presence of torsion 
the question of manifolds with a block-diagonal curvature
is an unsolved problem, except for the group manifolds discussed
above (where $\hat{R}^+=0$), and the new examples given in the
Conclusions. 

\section{6. CONCLUSIONS}

We have studied the general Virasoro construction
\be 
T=-\frac{L_{ij}}{\a'} \partial x^j \partial x^j + {\cal O}(\a'^0)
\ee
at one loop
in the operator algebra of the general non-linear sigma model,
where $L$ is a spin-two spacetime tensor field called the inverse
inertia tensor. The construction is summarized by a unified 
Einstein-Virasoro master equation which describes the covariant coupling of 
$L$ to the spacetime fields $G$, $B$ and $\Phi_a$, where $G$ and $B$
are the metric and antisymmetric tensor of the sigma model and
$\Phi_a$ is the dilaton vector, which generalizes the derivative
$\del_a \Phi$ of the dilaton $\Phi$.
As special cases, the unified system contains the Virasoro master equation
of the general affine-Virasoro construction 
and the conventional Einstein equations of the canonical sigma model
stress tensors.  More generally, the unified system describes a space of
 conformal field theories which is presumably much larger than the sum of
these two special cases.
    
In addition to questions posed in the text,
we list here a number of other important directions.

\noindent
1. New solutions. 
It is important to find new solutions of the unified system, beyond 
the canonical stress tensors of the sigma model 
and the general affine-Virasoro construction.
 
Although it is not in the original paper\refnote{\cite{dbh}}, we have recently
discovered a large class of new solutions of the covariant-constancy
condition: It is not hard to see
that the spin connection in the sigma model
description of the $g/h$ coset constructions has the form
\be
(\hat{\omega}^+_i)_a{}^b = N_i{}^A f_{Aa}{}^b
\ee
where $A$ is an $h$-index and $a,b$ are $g/h$-indices,
and $f_{Aa}{}^b$ are the structure constants of $g$. 
The structure constants and hence the spin connection can
be taken block diagonal, where the blocks correspond to
irreducible representations of $h$. As discussed in 
Section~5.5, this allows us to classify all possible
covariantly-constant projectors on these manifolds. 
More work remains to be done in this case, including
the solution of the generalized VME, but there are
indications that the resulting conformal field theories
may be identified as the set of local Lie $h$-invariant
conformal field theories\refnote{\cite{lie}} on $g$ , which have in fact
been studied in the Virasoro master equation itself.
 
\noindent
2. Duality. 
The unified system contains the coset constructions in two distinct 
ways, that is, both as $G_{ab}=k \eta_{ab}$, $L^{ab}=L^{ab}_{g/h}$
 in the general affine-Virasoro
construction and among the canonical stress tensors of
the sigma model with the sigma model metric that corresponds to the
coset construction. 
This is an indicator of new duality transformations in
the system, possibly exchanging $L$ and $G$, 
which may go beyond the coset constructions.
Indeed, if the conjecture of the previous paragraph holds, this
duality would extend over all local Lie $h$-invariant 
conformal field theory, and perhaps beyond. 
 
In this connection, we 
remind the reader that the VME has been identified\refnote{\cite{gme}}
 as an Einstein-Maxwell system with torsion on the group manifold, 
where the inverse inertia tensor is the inverse metric on
tangent space. Following this hint, it may be possible to cast the 
unified system on group manifolds as two coupled Einstein
systems, with exact covariant constancy of both $G$ and $L$.

\noindent
3. Non-renormalization theorems. 
The unified Einstein-Virasoro master equation is at present
a one-loop result, while the Virasoro master equation is
exact to all orders. This suggests a number of possibly 
exact relations\refnote{\cite{dbh}} to all orders in the WZW model and in
the general non-linear sigma model.

\noindent
4. Spacetime action and/or $C$-function. 
These have not yet been found for the unified system,
but we remark that they are known for the
special cases unified here: The spacetime action\refnote{\cite{sig9,ghor}}
is known for the conventional Einstein equations of the
sigma model, and, for this case,
the $C$-function is known\refnote{\cite{sig13}}
  for constant dilaton.
Moreover, an exact $C$-function is known\refnote{\cite{cf}}
for the special
case of the unimproved VME.

\noindent
5. World-sheet actions. We have studied here only the Virasoro 
operators constructible in the operator algebra of the general sigma model,
but we have not yet worked out the world-sheet actions of the 
corresponding new conformal field theories, whose beta functions
should be the unified Einstein-Virasoro master equation. 
This is a familiar situation in
the general affine-Virasoro construction, whose Virasoro operators 
are constructed in the operator algebra of the WZW model, while
the world-sheet actions of the corresponding new conformal field theories
include spin-one\refnote{\cite{br}} gauged WZW models for the coset 
constructions and spin-two\refnote{\cite{hy,ts3,bch}}
 gauged WZW models for the generic construction. 

As a consequence of 
this development in the general affine-Virasoro construction, 
more or less standard Hamiltonian
methods now exist for the systematic construction of the new world-sheet
actions from the new stress 
tensors, and we know for example that $K$-conjugation
covariance is the source of the spin-two gauge invariance in the generic case.
At least at one loop, a large subset of Class IIb solutions of the
unified system exhibit $K$-conjugation covariance,
so we may reasonably expect that the world-sheet
actions for generic constructions in this subset are
spin-two gauged sigma models. For solutions with no
$K$-conjugation covariance, the possibility remains open
that these constructions
 are dual descriptions of other conformal sigma models.

\noindent
6. Superconformal extensions. The method of Ref.~{\cite{sig12}}
has been extended\refnote{44--46} to the
canonical stress tensors of the supersymmetric
sigma model. The path is therefore open to study  general 
superconformal constructions in the operator algebra of the general
sigma model with fermions.
Such superconformal extensions should then include and 
generalize the known $N=1$ and $N=2$ superconformal master
equations\refnote{\cite{sme}} of the general affine-Virasoro construction. 

In this connection, we should mention that that the Virasoro master
equation is the true master equation, because it includes as a small 
subspace all the solutions of the superconformal master
equations. It is reasonable to expect therefore that, in the same way, 
the unified system of this paper will include the superconformal
extensions.

\section*{Acknowledgements} For many helpful discussions, we would
like to thank our colleagues:
L. Alvarez-Gaum\'e, I. Bakas,
K. Clubok, S. Deser,  A. Giveon, 
R. Khuri, E. Kiritsis, 
P. van Nieuwenhuizen,
N. Obers, A. Sen, 
K. Sfetsos, K. Skenderis,
A. Tseytlin and B. de Wit.
We also thank the organizers of the Zakopane Conference
for the opportunity to summarize this work.

This research was supported in part by the Director, Office of
Energy Research, Office of High Energy and Nuclear Physics, Division of
High Energy Physics of the U.S. Department of Energy under Contract
DOE-AC03-76SF00098 and in part by the National Science Foundation under
grant PHY-951497. JdB is a fellow of the Miller Institute for Basic
Research in Science.

\end{document}